# Exact solutions of the Schrödinger equation with non-central potential by Nikiforov-Uvarov method

F. Yaşuk[1], C. Berkdemir [2], A. Berkdemir [3]

Department of Physics, Faculty of Arts and Sciences, Erciyes University, 38039 Kayseri, Turkey

**Abstract**

The general solutions of Schrödinger equation for non-central potential are obtained by using Nikiforov-Uvarov method. The Schrödinger equation with general non-central potential is separated into radial and angular parts and energy eigenvalues and eigenfunctions for these potentials are derived analytically. Non-central potential is reduced to Coulomb and Hartmann potential by making special selections, and the obtained solutions are compared with the solutions of Coulomb and Hartmann ring-shaped potentials given in literature.

[1]e-mail: yasuk@erciyes.edu.tr
[2]e-mail: berkdemir@erciyes.edu.tr
[3]e-mail: arsland@erciyes.edu.tr

## 1. Introduction

One of the interesting problems of the nonrelativistic quantum mechanics is to find exact solutions to the Schrödinger equation for certain potentials of the physical interest. In recent years, considerable efforts have been done to obtain the analytical solution of non-central problems. In particular, the Coulombic ring-shaped potential [1] revived in quantum chemistry by Hartmann and coworkers [2] and the oscillatory ring-shaped potential, systematically studied by Quesne [3] have been investigated from a quantum mechanical view-point by using various approaches.

The Coluombic ring-shaped, or Hartmann, potential is,

$$V = -Z\frac{1}{\sqrt{x_1^2 + x_2^2 + x_3^2}} + \frac{1}{2}Q\frac{1}{x_1^2 + x_2^2}, \quad Z > 0, \quad Q > 0 \qquad (1)$$

where $Z = \eta\sigma^2$ and $Q = q\eta^2\sigma^2$ in the notation of Hartmann and of Kibler and Negadi [1]. This potential in the limiting case Q=0 reduces to an attractive Coulomb potential and a special case of the potential (in spherical coordinates)

$$V(r,\theta) = \frac{\alpha}{r} + \frac{\beta}{r^2 \sin^2\theta} + \gamma\frac{\cos\theta}{r^2 \sin^2\theta} \qquad (2)$$

introduced by Makarov et al. [4].

As for the solutions of Schrödinger equation for non-central potentials, it is seen that there are different methods used to obtain exact solutions which are supersymmetric [5, 6], path integral [7] and Bessel [8, 9].

In this letter, we introduce an alternative, elegant and simple method for an algebraic solution of the Schrödinger equation with non-central potential. This method called Nikiforov-Uvarov (NU) [10] is based on solving the second-order linear differential equations by reducing to a generalized equation of hypergeometric type.

NU-method is used to solve Schrödinger, Dirac, Klein-Gordon and Duffin-Kemmer-Petiau wave equations in the presence of exponential type potentials such as standard Woods-Saxon [11], Pöschl-Teller [12] and Hulthen [13,14]. The aim of this study is to illustrate that the Nikiforov-Uvarov method can be used to obtain exact solutions of non-central but separable potentials. Hence, radial and angular parts of Schrödinger equation with non-central potential is solved by NU-method and it is seen that this method is applicable to not only exponential but also non-central type potential.

This paper is arranged as follows: in section 2, the Schrödinger equation in spherical coordinates for a particle in the presence of non-central potential is separated into radial and angular parts. In section 3, Nikiforov-Uvarov method is given briefly . Then in section 4, the solutions of the radial and angular parts of the Schrödinger equation and the special cases of non-central, i.e., Hartmann and Coulomb potential's solution are obtained and compared with studies using different methods in the literature. Finally, the relevant results are discussed in section 5.

## 2. Separating variables of the Schrödinger equation with non-central potential

In the spherical coordinates, the Schrödinger equation with non-central potential is

$$-\frac{\hbar^2}{2m}\left[\frac{1}{r^2}\frac{\partial}{\partial r}\left(r^2\frac{\partial}{\partial r}\right)+\frac{1}{r^2\sin\theta}\frac{\partial}{\partial\theta}\left(\sin\theta\frac{\partial}{\partial\theta}\right)+\frac{1}{r^2\sin^2\theta}\frac{\partial^2}{\partial\varphi^2}\right]\psi(r,\theta,\varphi) \\ +\left(\frac{\alpha}{r}+\frac{\beta}{r^2\sin^2\theta}+\frac{\gamma\cos\theta}{r^2\sin^2\theta}\right)\psi(r,\theta,\varphi) = E\psi(r,\theta,\varphi) \qquad (3)$$

If we assign the spherical wave function as

$$\psi(r,\theta,\varphi) = R(r)Y(\theta,\varphi) \qquad (4)$$

we get

$$\frac{1}{R(r)}\frac{\partial}{\partial r}\left(r^2\frac{\partial R(r)}{\partial r}\right)-\frac{2m\alpha r}{\hbar^2}+\frac{2mEr^2}{\hbar^2}+\frac{1}{Y(\theta,\varphi)}\frac{1}{\sin\theta}\frac{\partial}{\partial\theta}\left(\sin\theta\frac{\partial Y(\theta,\varphi)}{\partial\theta}\right) \\ -\frac{2m}{\hbar^2}\left(\frac{\beta}{\sin^2\theta}+\frac{\gamma\cos\theta}{\sin^2\theta}\right)+\frac{1}{Y(\theta,\varphi)}\frac{1}{\sin^2\theta}\frac{\partial^2 Y(\theta,\varphi)}{\partial\varphi^2}=0 \qquad (5)$$

Separating the Schrödinger equation into variables and selecting $Y(\theta,\varphi) = H(\theta)\Phi(\varphi)$, the following equations are obtained;

$$\frac{d^2R(r)}{dr^2}+\frac{2}{r}\frac{dR(r)}{dr}+\frac{2mr^2}{\hbar^2}\left(E-\frac{\alpha}{r}-\frac{\lambda}{r^2}\right)R(r)=0, \qquad (6a)$$

$$\frac{d^2H(\theta)}{d\theta^2}+\cot\theta\frac{dH(\theta)}{d\theta}+\left(\lambda-\frac{m^2}{\sin^2\theta}-\frac{2m}{\hbar^2}\left(\frac{\beta+\gamma\cos\theta}{\sin^2\theta}\right)\right)H(\theta)=0 \qquad (6b)$$

$$\frac{d^2\Phi(\varphi)}{d\varphi^2} + m^2\Phi(\varphi) = 0 \tag{6c}$$

where $m^2$ and $\lambda$ are separation constants. It is well-known that the solutions of the equation (6c) is,

$$\Phi_m = Ae^{im\varphi}, \qquad (m=0, \pm1, \pm2\ldots) \tag{7}$$

Eq. (6a) and (6b) are radial and polar angle equations and they will be solved by using Nikiforov-Uvarov method, given briefly in the following section.

## 3. Nikiforov-Uvarov Method

In this method, for a given real or complex potential, the Schrödinger equation in one dimension is reduced to a generalized equation of hypergeometric type with an appropriate $s = s(r)$ coordinate transformation and it can be written in the following form,

$$\psi(s)'' + \frac{\tilde{\tau}(s)}{\sigma(s)}\psi'(s) + \frac{\tilde{\sigma}(s)}{\sigma^2(s)}\psi(s) = 0 \tag{8}$$

where $\sigma(s)$ and $\tilde{\sigma}(s)$ are polynomials, at most second-degree, and $\tilde{\tau}(s)$ is a first-degree polynomial. Hence, from the Eq. (8) the Schrödinger equation and the Schrödinger-like equations can be solved by means of the special potentials or some quantum mechanics problems. To find particular solution of Eq. (8) by separation of variables, if one deals with the transformation

$$\psi(s) = \phi(s)y(s) \tag{9}$$

it reduces to an equation of hypergeometric type,

$$\sigma(s)y'' + \tau(s)y' + \lambda y = 0 \tag{10}$$

and $\phi(s)$ is defined as a logaritmic derivative

$$\phi'(s)/\phi(s) = \pi(s)/\sigma(s) \tag{11}$$

The other part $y(s)$ is the hypergeometric type function whose polynomial solutions are given by Rodrigues relation

$$y_n(s) = \frac{B_n}{\rho(s)} \frac{d^n}{ds^n}\left[\sigma^n(s)\rho(s)\right] \tag{12}$$

where $B_n$ is a normalizing constant and the weight function $\rho(s)$ must satisfy the condition

$$(\sigma\rho)' = \tau\rho \tag{13}$$

The function $\pi$ and the parameter $\lambda$ required for this method are defined as follows

$$\pi(s) = \frac{\sigma' - \tilde{\tau}}{2} \pm \sqrt{\left(\frac{\sigma' - \tilde{\tau}}{2}\right)^2 - \tilde{\sigma} + k\sigma}, \tag{14}$$

$$\lambda = k + \pi' \tag{15}$$

On the other hand, in order to find the value of k, the expression under the square root must be square of polynomial. Thus, a new eigenvalue equation for the Schrödinger equation becomes

$$\lambda = \lambda_n = -n\tau' - \frac{n(n-1)}{2}\sigma'' \tag{16}$$

where

$$\tau(s) = \tilde{\tau}(s) + 2\pi(s) \tag{17}$$

and its derivative is negative.

## 4. Solutions of the Radial and Polar Angle equations with NU Method

### 4.1. Eigenvalues and Eigenfunctions of the Radial Equation

We assume that $R(r) = (1/r)F(r)$ is bounded as $r \to 0$, radial Schrödinger equation given in Eq.(6a) is

$$F''(r) + \left(\frac{2mE}{\hbar^2} - \frac{2m\alpha}{\hbar^2 r} - \frac{\ell(\ell+1)}{r^2}\right)F(r) = 0 \tag{18}$$

Letting

$$\frac{2mE}{\hbar^2} = -\varepsilon^2 \quad ; \quad \frac{2m\alpha}{\hbar^2} = b^2 \quad ; \quad \lambda = \ell(\ell+1) \quad ; \quad \alpha = -Ze^2. \tag{19}$$

and substituting these expressions in Eq.(18), one obtains as follows:

$$F''(r) + \left(-\varepsilon^2 r^2 - b^2 r - \lambda\right)\frac{1}{r^2} F(r) = 0 \tag{20}$$

To apply the NU method by comparing Eq.(20) with Eq.(8), we get

$$\tilde{\tau} = 0 \quad ; \quad \sigma = r \quad ; \quad \tilde{\sigma} = -\varepsilon^2 r^2 - b^2 r - \lambda \tag{21}$$

Inserting these polynomials in Eq.(14), we achive $\pi$ function as

$$\pi = \frac{1}{2} \pm \frac{1}{2}\sqrt{4\varepsilon^2 r^2 + 4r(k+b^2) + 4\lambda + 1} \tag{22}$$

According to the NU method, the expression in the square root must be square of polynomial. So, one can find new possible functions for each k as

$$\pi = \begin{cases} \frac{1}{2} \pm \left[\sqrt{\varepsilon^2} r + \left(\ell + \frac{1}{2}\right)\right] & , \text{for} \quad k = -b^2 + 2\sqrt{\varepsilon^2}\left(\ell + \frac{1}{2}\right) \\ \frac{1}{2} \pm \left[\sqrt{\varepsilon^2} r - \left(\ell + \frac{1}{2}\right)\right] & , \text{for} \quad k = -b^2 - 2\sqrt{\varepsilon^2}\left(\ell + \frac{1}{2}\right) \end{cases} \tag{23}$$

For the polynomial of $\tau = \tilde{\tau} + 2\pi$ which has a negative derivative, we select

$$k = -b^2 - 2\sqrt{\varepsilon^2}\left(\ell + \frac{1}{2}\right) \quad \text{and} \quad \pi = \frac{1}{2} - \left[\sqrt{\varepsilon^2} r - \left(\ell + \frac{1}{2}\right)\right] \tag{24}$$

With this selection, and $\lambda = k + \pi'$, $\tau$ and $\lambda$ can be written as, respectively,

$$\tau = 2\left(\ell + 1 - \sqrt{\varepsilon^2} r\right) \tag{25}$$

$$\lambda = -b^2 - \sqrt{\varepsilon^2}(2\ell + 2) \tag{26}$$

Another definition of $\lambda_N$ at Eq.(16),

$$\lambda_N = 2N\sqrt{\varepsilon^2} \tag{27}$$

and comparing with Eq.(26), the exact energy eigenvalues of radial part of Schrödinger equation with non-central potential are derived as,

$$E_N = -\frac{mZ^2 e^4}{\hbar^2} \frac{1}{2(N+\ell+1)^2} \tag{28}$$

Having obtained energy eigenvalues of radial Schrödinger equation with non-central potential, now let us consider the corresponding wave functions. Using $\sigma$ and $\pi$ in Eqs.(11-13), one can find wave functions $y(r) = y_{N\ell}(r)$ and $\phi(r)$ from Eq.(9)

$$F_{N\ell}(z) = C_{N\ell} z^{\ell+1} \exp(-\frac{z}{2}) L_N^{2\ell+1}(z) \tag{29}$$

where $L_N^{2\ell+1}(z)$ stands for the associated Laguerre functions whose argument is equal to $z = \frac{2\mu Z e^2}{\hbar^2(N+\ell+1)} r$ and $C_{N\ell}$ is normalization constant determined by $\int_0^\infty F_{N\ell}^2(r) dr = 1$ [15]. Thus, the corresponding normalized wave functions are found to be

$$F_{n'\ell}(r) = \left(\frac{\mu Z e^2}{\hbar^2 n'}\right)^{1/2} \left(\frac{(n'-\ell-1)!}{n'\Gamma(n'+\ell+1)}\right)^{1/2} \left(\frac{2\mu Z e^2}{\hbar^2 n'}\right)^{\ell+1} r^{\ell+1} \exp\left(-\frac{2\mu Z e^2 r}{\hbar^2 n'}\right) L_{n'-\ell-1}^{2\ell+1}\left(\frac{2\mu Z e^2}{\hbar^2 n'} r\right) \tag{30}$$

where $n' = N + \ell + 1$. This equation is also stands for solution of radial Schrödinger equation with Coulomb potential, since radial Schrödinger equation with non-central potential contains only Coulombic potential terms.

**4.2. Eigenvalues and Eigenfunctions of the Polar Angle Equation**

We are now going to derive eigenvalues and eigenfunctions of polar angle part of Schrödinger equation with similar method as given in subsection 4.1.

Introducing a new variable $\cos\theta = x$ and then Eq.(6b) is brought to the form,

$$\frac{d^2 H(x)}{dx^2} - \frac{2x}{1-x^2} \frac{dH(x)}{dx} + \left(\frac{\lambda(1-x^2) - m^2 - \frac{2m}{\hbar^2}(\beta + \gamma x)}{(1-x^2)^2}\right) H(x) = 0 \tag{31}$$

Comparing with Eq.(8), the following expressions are obtained

$$\tilde{\tau} = -2x \quad ; \quad \sigma = 1 - x^2 \quad ; \quad \tilde{\sigma} = -\lambda x^2 - \gamma x + (\lambda - m^2 - \beta) \tag{32}$$

Putting them in Eq.(14), the function $\pi$ is,

$$\pi = \pm\sqrt{x^2(\lambda - k) + \gamma x - (\lambda - m^2 - \beta - k)} \tag{33}$$

According to the NU method, the expression in the square root must be square of polynomial. So, one can find new possible functions for each k as

$$\pi = \pm \begin{cases} x\sqrt{\dfrac{m^2+\beta+u}{2}} + \sqrt{\dfrac{m^2+\beta-u}{2}} & ,\text{ for } k = \dfrac{2\lambda - m^2 - \beta}{2} - \dfrac{1}{2}u \\ x\sqrt{\dfrac{m^2+\beta-u}{2}} + \sqrt{\dfrac{m^2+\beta+u}{2}} & ,\text{ for } k = \dfrac{2\lambda - m^2 - \beta}{2} + \dfrac{1}{2}u \end{cases} \tag{34}$$

where $u = \sqrt{(m^2+\beta)^2 - \gamma^2}$.

For the polynomial of $\tau = \tilde{\tau} + 2\pi$ which has a negative derivative,

$$\tau = -2\sqrt{\dfrac{m^2+\beta-u}{2}} - 2x\left(1 + \sqrt{\dfrac{m^2+\beta+u}{2}}\right) \tag{35}$$

$\lambda = k + \pi'$ and another definition $\lambda_n = -n\tau' - \dfrac{n(n-1)}{2}\sigma''$ are given as follows, respectively,

$$\lambda = \dfrac{2\lambda - (m^2+\beta)}{2} - \dfrac{1}{2}u - \sqrt{\dfrac{m^2+\beta+u}{2}} \tag{36}$$

$$\lambda_n = 2n\left(1 + \sqrt{\dfrac{m^2+\beta+u}{2}}\right) + n(n-1) \tag{37}$$

To obtain $E_n = \lambda - (m^2+\beta)$, we compared Eq.(36) and Eq.(37),

$$(2n+1)\sqrt{\dfrac{m^2+\beta+u}{2}} + \dfrac{u-(m^2+\beta)}{2} + n(n+1) = \lambda - (m^2+\beta) \tag{38}$$

Using the definition of $\lambda = \ell(\ell+1)$, from Eq.(38) one obtains

$$\ell = \sqrt{\dfrac{m^2+\beta+\sqrt{(m^2+\beta)^2-\gamma^2}}{2}} + n \tag{39}$$

If we substitute Eq.(39) into eigenvalues of radial part of Schrödinger equation with non-central potential, which has similar radial Schrödinger equation to Coulomb potential, Eq.(28), we found the final energy eigenvalues for a bound electron in a Coulomb potential as well a combination of non-central potentials is given by Eq.(6b) are

$$E_N = -\frac{mZ^2e^4}{\hbar^2} \frac{1}{2\left(N + \sqrt{\frac{m^2 + \beta + \sqrt{(m^2+\beta)^2 - \gamma^2}}{2}} + n + 1\right)^2} \tag{40}$$

As for wave function of polar angle part of Schrödinger equation, using σ and π in Eqs.(11-13), one obtains

$$\phi = (1-x)^{B+C/2}(1+x)^{B-C/2} \tag{41}$$

$$\rho = (1-x^2)^B \left(\frac{1+x}{1-x}\right)^{-C} \tag{42}$$

$$y_n = B_n(1-x)^{-(B+C)}(1+x)^{-(B-C)} \frac{d^n}{dx^n}\left[(1+x)^{n+B-C}(1-x)^{n+B+C}\right] \tag{43}$$

where $B = \sqrt{\frac{m^2+\beta+u}{2}}$ and $C = \sqrt{\frac{m^2+\beta-u}{2}}$. The polynomial solution of $y_n$ is expressed in terms of Jacobi polynomials which is one of the ortogonal polynomials, giving $\approx P_n^{(B+C, B-C)}(x)$. Substituting Eqs.(41, 43) into Eq.(9), the correspnding wave functions are found to be

$$H_n(x) = N_n (1-x)^{(B+C)/2}(1+x)^{(B-C)/2} P_n^{(B+C, B-C)}(x) \tag{44}$$

where $N_n$ is normalization constant determined by $\int_{-1}^{+1}[H_n(x)]^2 dx = 1$ and using the relation orthogonality of Jacobi polynomials [15,16], the normalization constant becomes

$$N_n = \sqrt{\frac{(2n+2B+1)\Gamma(n+1)\Gamma(n+2B+1)}{2^{2B+1}\Gamma(n+B+C+1)\Gamma(n+B-C+1)}} \tag{45}$$

### 4.3. A Special Case: Hartmann ring-shaped potential

The ring-shaped Hartmann potential is

$$V_H(r,\theta) = \eta\sigma^2\varepsilon_0\left(\frac{2a}{r} - \frac{q\eta a^2}{r^2 \sin^2\theta}\right) \tag{46}$$

Here, $a = \hbar^2/\mu e^2$ (Bohr's radius) and $\varepsilon_0 = -\mu e^4/2\hbar^2$ (ground-state energy of the hydrogen atom), $\eta$ and $\sigma$ are dimensionless positive parameters which range from about 1 to 10 in theoretical chemistry applications and q is a real parameter. If we compare the Hartmann potential with generalized non-central potential given in Eq.(2), the following expressions are found to be $\alpha = -\eta\sigma^2 e^2$, $\beta = \dfrac{q\eta^2\sigma^2\hbar^2}{2\mu}$ and $\gamma = 0$. Using the similarity of separated equations for the Hartmann potential with those of the hydrogen atom, one can write immediately from Eq.(40), the energy spectrum for the Hartmann system is given by

$$E_N = -\frac{\mu(\eta\sigma^2)^2 e^4}{2\hbar^2\left(N + \sqrt{m^2 + \dfrac{q\eta^2\sigma^2\hbar^2}{2\mu}} + n + 1\right)^2} \tag{47}$$

and radial wave functions of Hartmann potential is given in Eq.(30) and in agreement with [5,8].

Now let us obtain polar angle wave functions of Hartmann potential by reducing polar angle wave functions of generalized non-central potential as given in Eq.(44). When $\gamma = 0$, we simply get the parameters as $u = m^2 + \beta$, $B = (m^2 + \beta)^{1/2}$ and $C = 0$. With the selection of $m' = B$, we get

$$H_n(\cos\theta) = \sqrt{\frac{(2n + 2m' + 1)\Gamma(n+1)\Gamma(n + 2m' + 1)}{2^{2m'+1}\Gamma(n + m' + 1)\Gamma(n + m' + 1)}}(\sin\theta)^{m'} P_n^{(m',m')}(\cos\theta) \tag{48}$$

Here, $P_n^{(m',m')}(\cos\theta)$ is given in terms of ultraspherical polynomials as

$$P_n^{(m',m')}(\cos\theta) = \frac{\Gamma(2m'+1)}{\Gamma(m'+1)} \frac{\Gamma(n + m' + 1)}{\Gamma(n + 2m' + 1)} P_n^{(\lambda)}(\cos\theta) \tag{49}$$

and using the definition of ultraspherical polynomials we get Eq.(48) as follows

$$H_{\ell m'}(\cos\theta) = \sqrt{\frac{(2n + 2m' + 1)n!}{2\Gamma(n + 2m' + 1)}}(\sin\theta)^{m'} \sum_{v=0}^{\left[\frac{\ell - m'}{2}\right]} \frac{(-1)^v \Gamma(2\ell - 2v + 1)}{2^\ell v!(\ell - m' - 2v)!\Gamma(\ell - v + 1)}(\cos\theta)^{(\ell - m' - 2v)} \tag{50}$$

where $\ell = n + m'$, n=0, 1, 2, .... Therefore, we derived polar angle wave functions of Hartmann potential by reducing polar angle wave functions of generalized non-central potential and Eq.(50) is consistent with [8].

## 5. Conclusions

We have obtained the exact solutions of the radial and angular parts of Schrödinger equation for non-central potential using Nikiforov-Uvarov method. This method is usually used in solving analytically Schrödinger, Dirac, Klein-Gordon and Duffin-Kemmer-Petiau wave equations in the presence of exponential type potentials. In this study, we applied NU-method to non-central potentials by separating SE into radial and spherical polar coordinates, so we generalized the feasibility of the NU-method. Energy eigenvalues are obtained in a Coulomb potential as well a combination of non-central potentials. This calculation has been done by the path integral solution of the system with/without using Kustannheimo-Stiefel transformation in [7]. Energy eigenfunctions are derived for radial and polar angle parts of Schrödinger equation with non-central potential and radial and polar angle wave functions are found in terms of Laguerre and Jacobi polynomials, respectively. When $\alpha = -\eta\sigma^2 e^2$, $Z = \eta\sigma^2$, $\beta = \dfrac{q\eta^2\sigma^2\hbar^2}{2\mu}$ and $\gamma = 0$, non-central potential reduces to Hartmann potential. The energy spectrum of the Hartmann system is obtained as in [5,8] and polar angle wavefunctions are found in terms of ultraspherical Jacobi polynomials. The relevant results of Schrödinger equation for not only non-central but also special cases of non-central, i.e, Hartmann and Coulomb potentials are compared with studies using different methods in the literature.

**Acknowledgement**

The authors wish to thank Professor Coşkun Önem for many helpful discussions and suggested improvements to the paper.